\documentclass[11pt,onecolumn,draftcls]{IEEEtran}

\usepackage{graphicx}
\usepackage{mathptmx}
\usepackage{amsmath,bm}
\usepackage{latexsym}

\begin{document}

\title{On the efficient estimation of the mean of multivariate truncated normal distributions\thanks{This research has been co-financed by the European Union (European Social Fund - ESF) and Greek national funds through the Operational Program ``Education and Lifelong Learning'' of the National Strategic Reference Framework (NSRF) - Research Funding Program: ARISTEIA- HSI-MARS-1413.}
}

\author{{ Konstantinos D. Koutroumbas}\thanks{Institute for Astronomy, Astrophysics,
Space Applications and Remote sensing, National Observatory of Athens, I. Metaxa
and Vas. Paulou, Penteli, GR-152 36, Greece, Tel. No: +30-210-8109189,
Fax No: +30-210-6138343, E-mail: koutroum@noa.gr.}  
\and { Konstantinos E. Themelis}
\and { Athanasios A. Rontogiannis}
 }

\date{23 January 2014}

\maketitle

\begin{abstract}
A non trivial problem that arises in several applications is the estimation of the mean of a truncated normal distribution. In this paper, an iterative deterministic scheme for approximating this mean is proposed, motivated by an iterative Markov chain Monte Carlo (MCMC) scheme that addresses the same problem. Conditions are provided under which it is proved that the scheme converges to a unique fixed point. 
The quality of the approximation obtained by the proposed scheme is assessed through the case where the exponential correlation matrix is used as covariance matrix of the initial (non truncated) normal distribution. 
Finally, the theoretical results are also supported by computer simulations, which show the rapid convergence of the method to a solution vector that (under certain conditions) is very close to the mean of the truncated normal distribution under study.
\end{abstract}


\section{Introduction}
\label{intro}
A non trivial problem that appears in several applications, e.g. in multivariate regression and Bayesian statistics, is the estimation of the mean of a truncated normal distribution. The problem arises in cases where a random vector ${\mbox{\boldmath $x$}}=[x_1,\ldots,x_n]^T$ follows a normal distribution with mean ${\mbox{\boldmath $\mu$}}$ and covariance matrix $\Sigma$, denoted by ${\cal N}({\mbox{\boldmath $\mu$}}, \Sigma)$, but ${\mbox{\boldmath $x$}}$ is restricted to a closed subset $R^n$ of ${\cal R}^n$. However, the restrictions considered in most of the cases are of the form $a_i \leq x_i \leq b_i$, with $a_i,b_i \in {\cal R} \cup \{-\infty, + \infty \}$, $i=1,\ldots,n$. If $a_i \in {\cal R}$ and $b_i=+\infty$ ($a_i=-\infty $ and $b_i \in {\cal R}$), we have a {\em one-sided truncation} from the left (right), while in the case where both $a_i$ and $b_i$ are reals, we have {\em two-sided truncation}. 

The problem has attracted the attention of several researchers from the sixties. Since then, several deterministic approaches have been proposed with some of them trying to estimate not only the mean but also other moments (such as the variance) of a multivariate truncated normal distribution. These approaches can be categorized according to whether they are suitable for one-sided truncated normal distributions (\cite{Rose61}, \cite{Tall61}, \cite{Amem74}, \cite{Lee79}, \cite{Lepp89}) or for two-sided truncated normal distributions (\cite{Shah64}, \cite{Regi71}, \cite{Gupt76}, \cite{Lee83}, \cite{Muth90}, \cite{Manj12}), or according to whether they consider the bivariate, $n=2$, (\cite{Rose61}, \cite{Shah64}, \cite{Regi71}, \cite{Muth90}) or the multivariate case (\cite{Tall61}, \cite{Amem74}, \cite{Gupt76}, \cite{Lee79}, \cite{Lee83}, \cite{Lepp89}, \cite{Manj12}). In addition, some of these methods put additional restrictions to the distribution paramaters (e.g. \cite{Tall61}, \cite{Amem74}, \cite{Lee79} require that ${\mbox{\boldmath $\mu$}}={\mbox{\boldmath $0$}}$).
Most of these methods either perform direct integration (e.g. \cite{Tall61}) or they utilize the moment generating function tool (e.g. \cite{Manj12}).

An alternative strategy to deal with this problem has been followed in \cite{Robe95}. More specifically, in \cite{Robe95}, a Markov Chain Monte Carlo (MCMC) iterative scheme has been developed. According to this, at each iteration, $n$ succesive samplings take place, one from each one-dimensional conditionals of the truncated normal distribution. The mean of the $i$-th such distribution is the mean of $x_i$ conditioned on the $x_1,\ldots,x_{i-1},x_{i+1},\ldots,x_n$. After performing several iterations, the estimation of the mean of the truncated normal results by performing an averaging over the produced samples. Convergence issues of this scheme to the mean of the truncated normal distribution are a subject of the Markov chain theory. A relative work that accelerates the method in \cite{Robe95} is exhibited in \cite{Chop11}.  

The work presented in this paper for approximating the mean of a multivariate truncated normal distribution has been inspired from that of \cite{Robe95}. Specifically, instead of selecting a sample from each one of the above one-dimensional distributions, we select its mean. Thus, the proposed scheme departs from the statistical framework adopted in \cite{Robe95} and moves to the deterministic one.

This work is an extension of a relative scheme used in \cite{Them12} in the framework of spectral unmixing in hyperspectral images. In addition, a convergence proof of the proposed scheme is given when certain conditions are fulfilled. The quality of the approximation of the mean offered by the proposed method is assessed via the case where $\Sigma$ is the exponential correlation matrix. Experimental results show that the new scheme converges significantly faster than the MCMC approach of \cite{Robe95}.

The paper is organized as follows. Section 2 contains some necessary definitions and a brief description of the work in \cite{Robe95}. In Section 3, the newly proposed method is described and in Section 4 conditions are given under which it is proved to converge. In Section 5, the proposed method is applied to the case where $\Sigma$ is the exponential correlation matrix. In Section 6 simulation results are provided and a relevant discussion is presented. Finally, Section 7 concludes the paper.

\section{Preliminaries and previous work}
\label{section2}
Let us consider the $n$-dimensional normal distribution
\begin{equation}
{\cal N}({\mbox{\boldmath $x$}}|{\mbox{\boldmath $\mu$}}, \Sigma) = \frac{1}{(2\pi)^{n/2} |\Sigma|^{1/2}} \exp(-\frac{1}{2} ({\mbox{\boldmath $x$}}-{\mbox{\boldmath $\mu$}})^T \Sigma^{-1} ({\mbox{\boldmath $x$}}-{\mbox{\boldmath $\mu$}}) )
\end{equation}
where the $n$-dimensional vector ${\mbox{\boldmath $\mu$}}$ is its mean and the $n \times n$ matrix $\Sigma$ is its covariance matrix.

Let $R^n$ be a subset of ${\cal R}^n$ with positive Lebesgue measure. We denote by ${\cal N}_{R^n}({\mbox{\boldmath $x$}}|{\mbox{\boldmath $\mu$}}, \Sigma)$ the truncated normal distribution which results from the truncation of ${\cal N}({\mbox{\boldmath $x$}}|{\mbox{\boldmath $\mu$}}, \Sigma)$ in $R^n$. Speaking in mathematical terms
\begin{equation}
\label{trunc}
{\cal N}_{R^n}({\mbox{\boldmath $x$}}|{\mbox{\boldmath $\mu$}}, \Sigma) = \left \{
\begin{array}{cc}
\frac{\exp(-\frac{1}{2} ({\mbox{\boldmath $x$}}-{\mbox{\boldmath $\mu$}})^T \Sigma^{-1} ({\mbox{\boldmath $x$}}-{\mbox{\boldmath $\mu$}}) )}{\int_{R^n} \exp(-\frac{1}{2} ({\mbox{\boldmath $x$}}-{\mbox{\boldmath $\mu$}})^T \Sigma^{-1} ({\mbox{\boldmath $x$}}-{\mbox{\boldmath $\mu$}}) )d{\mbox{\boldmath $w$}} }, &  if\ {\mbox{\boldmath $x$}} \in R^n  \\
0, &  otherwise
\end{array} \right.
\end{equation}

Note that ${\cal N}_{R^n}({\mbox{\boldmath $x$}}|{\mbox{\boldmath $\mu$}}, \Sigma)$ is {\em proportional} to ${\cal N}({\mbox{\boldmath $x$}}|{\mbox{\boldmath $\mu$}}, \Sigma) I_{R^n}({\mbox{\boldmath $x$}})$, where $I_{R^n}({\mbox{\boldmath $x$}})=1$, if ${\mbox{\boldmath $x$}} \in R^n$ and $0$, otherwise.

In the scheme discussed in \cite{Robe95}, a Markov Chain Monte Carlo (MCMC) method is proposed, to compute the mean of single or doubly truncated (per coordinate) normal distribution ${\cal N}_{R^n}({\mbox{\boldmath $x$}}|{\mbox{\boldmath $\mu$}}, \Sigma)$, where $R^n= [a_1,b_1] \times [a_2,b_2] \times \ldots \times [a_n,b_n]$. The method relies on the sampling of the $n$ one-dimensional conditionals of the truncated normal distribution. More specifically, letting $\mu_i^* \equiv E[x_i|x_1,\ldots,x_{i-1},x_{i+1},\ldots,$ $x_n]$ and $\sigma_i^{*2}$ denoting the expectation and the variance of $x_i$ conditioned on the rest coordinates, respectively\footnote{That is, $x_i$ follows the (non-truncated) $i$-th conditional of ${\cal N}({\mbox{\boldmath $x$}}|{\mbox{\boldmath $\mu$}}, \Sigma)$.}, and ${\cal N}_{[a_i,b_i]}(\mu_i^*,$ $\sigma_i^{*2})$ denoting the (one dimensional) truncated normal distribution which results from the truncation of a normal distribution with mean $\mu_i^*$ and variance $\sigma_i^{*2}$ in $[a_i,b_i]$, the iterative sampling scheme proposed in \cite{Robe95} can be written as
\begin{equation}
\label{robert}
\begin{array}{ll}
1. & x_1^{(t)} \sim {\cal N}_{[a_1,b_1]}(E[x_1|x_2^{(t-1)},\ldots,x_n^{(t-1)}],\sigma_1^{*2}) \\
2. & x_2^{(t)} \sim {\cal N}_{[a_2,b_2]}(E[x_2|x_1^{(t)},x_3^{(t-1)}\ldots,x_n^{(t-1)}],\sigma_2^{*2}) \\
\vdots &  \\
n. & x_n^{(t)} \sim {\cal N}_{[a_n,b_n]}(E[x_n|x_1^{(t)},x_2^{(t)}\ldots, x_{n-1}^{(t)}],\sigma_n^{*2})
\end{array}
\end{equation}
where $\sim$ denotes the sampling action and $t$ denotes the current iteration. After performing several, say $k$, iterations (and after discarding the first few, say $k'$, ones) the mean of each coordinate is estimated as 
$$E[x_i]=\frac{1}{k-k'} \sum_{t=k'}^k x_i^{(t)}, \ \ i=1,\ldots,n$$

The quantities $\mu_i^*$ and $\sigma_i^*$ of each one of the above one dimensional conditionals are expressed in terms of the parameters ${\mbox{\boldmath $\mu$}}$ and $\Sigma$ of the non-truncated multidimensional normal distribution as follows
\begin{equation}
\label{trunc-mean}
\mu_i^*=\mu_i + {\mbox{\boldmath $\sigma$}}_{\neg i}^T \Sigma_{\neg i \neg i}^{-1} ({\mbox{\boldmath $x$}}_{\neg i} - {\mbox{\boldmath $\mu$}}_{\neg i}), \ \ \ i=1,\ldots,n
\end{equation}

\begin{equation}
\label{trunc-var}
\sigma_{i}^*=\sigma_{ii} - {\mbox{\boldmath $\sigma$}}_{\neg i}^T \Sigma_{\neg i \neg i}^{-1} {\mbox{\boldmath $\sigma$}}_{\neg i}, \ \ \ i=1,\ldots,n
\end{equation}

\noindent with $\Sigma_{\neg i \neg i}$ being the $(n-1)\times(n-1)$ matrix that results from $\Sigma$ after removing its $i$-th column and its $i$-th row, ${\mbox{\boldmath $\sigma$}}_{\neg i}$ being the $i$-th column of $\Sigma$ excluding its $i$-th element and ${\mbox{\boldmath $x$}}_{\neg i}$, ${\mbox{\boldmath $\mu$}}_{\neg i}$ being the ($(n-1)$-dimensional) vectors that result from ${\mbox{\boldmath $x$}}$ and ${\mbox{\boldmath $\mu$}}$, respectively, after removing their $i$-th coordinates, $x_i$ and $\mu_i$, respectively. Note that $\mu_i^*$ depends on all $x_j$'s {\em except} $x_i$.

\section{The proposed model}

In the sequel, we focus on the case where $R^n$ is a set of the form $[a_1,b_1] \times [a_2,b_2] \times \ldots \times [a_n,b_n]$, where for each interval $[a_i,b_i]$ it is either, (i) $a_i=-\infty$ and $b_i \in R$ or (ii) $a_i \in R$ and $b_i=\infty$. This means that, along each dimension, the truncation is one-sided and more specifically, case (i) corresponds to right truncation while case (ii) corresponds to left truncation. 

The proposed model for estimating the mean of ${\cal N}_{R^n}({\mbox{\boldmath $x$}}|{\mbox{\boldmath $\mu$}},$ $\Sigma)$, is of iterative nature and, at each iteration, it requires the computation of the (one-dimensional) $erfc$ function. 
This model has a close conceptual affinity with the one (briefly) presented in the previous section (\cite{Robe95}). More specifically, instead of utilizing the samples produced by the (one dimensional) distributions
${\cal N}_{[a_i,b_i]}(E[x_i|x_1^{(t)},x_2^{(t)} \ldots,x_{i-1}^{(t)}, x_{i+1}^{(t-1)},\ldots,$ $x_n^{(t-1)}],\sigma_i^{*2})$,
we use the corresponding mean values (here denoted by ${\mbox{\boldmath $w$}}$ $=[w_1,w_2,\ldots,w_n]^T$).

As it is well known (see e.g. \cite{Oliv98}), the mean $w$ of a truncated one dimensional normal distribution ${\cal N}_{[a,b]} (\mu^*,\sigma^{*2})$, which has resulted from the (non-truncated) normal distribution with mean $\mu^*$ and variance $\sigma^{*2}$, is expressed as
\begin{equation}
\label{truncmean1d}
w = \mu^* + \frac{\phi(\frac{a-\mu^*}{\sigma^*})-\phi(\frac{b-\mu^*}{\sigma^*})}{\Phi(\frac{b-\mu^*}{\sigma^*})-\Phi(\frac{a-\mu^*}{\sigma^*})} \sigma^*
\end{equation}
where $\phi(x)=\frac{1}{\sqrt{2\pi}} e^{-\frac{x^2}{2}}$, $\Phi(x)=\frac{1}{2}erfc(-\frac{x}{\sqrt{2}})$ and $erfc(x)=\frac{2}{\sqrt{\pi}} \int_x^{+\infty} e^{-t^2}dt$.

However, since in the present paper we consider the cases where either (i) $a_i \in R$ and $b_i=\infty$ or (ii) $a_i=-\infty$ and $b_i \in R$, let us see now how eq. (\ref{truncmean1d}) becomes for each of these cases.

(i) $a \in R$, $b=+\infty$. In this case it is $\frac{b-\mu^*}{\sigma^*}=+\infty$ and as a consequence, $\phi(\frac{b-\mu^*}{\sigma^*})=0$, $\Phi(\frac{b-\mu^*}{\sigma^*})=\frac{1}{2} erfc(-\frac{b-\mu^*}{\sqrt{2}\sigma^*})=\frac{1}{2} erfc(-\infty)=1$. Thus, taking also into account the definitions of $\phi$ and $\Phi$ and the fact that $erfc(x)+erfc(-x)=2$,  eq. (\ref{truncmean1d}) gives,
\begin{equation}
\label{case-i}
w=\mu^* + \sqrt{\frac{2}{\pi}} \frac{e^{-\frac{(\mu^*-a)^2}{2\sigma^{*2}}}}{erfc(-\frac{\mu^*-a}{\sqrt{2} \sigma^*})} \sigma^*
\end{equation} 

(ii) $a=-\infty$, $b \in R$. In this case it is $\frac{a-\mu^*}{\sigma^*}=-\infty$ and, as a consequence,
$\phi(\frac{a-\mu^*}{\sigma^*})=0$, $\Phi(\frac{a-\mu^*}{\sigma^*})=\frac{1}{2} erfc(-\frac{a-\mu^*}{\sqrt{2}\sigma^*})=\frac{1}{2} erfc(\infty)=0$. Working as in case (i), eq. (\ref{truncmean1d}) gives
\begin{equation}
\label{case-ii}
w=\mu^* - \sqrt{\frac{2}{\pi}} \frac{e^{-\frac{(b-\mu^*)^2}{2\sigma^{*2}}}}{erfc(-\frac{b-\mu^*}{\sqrt{2} \sigma^*})} \sigma^*
\end{equation} 

Eqs. (\ref{case-i}) and (\ref{case-ii}) can be expressed compactly via the following single equation
\begin{equation}
\label{cases-i-ii}
w=\mu^* + \sqrt{\frac{2}{\pi}} \left [ f(\frac{\mu^*-a}{\sqrt{2} \sigma^*}) I_{a \in {\cal R}} - 
f(\frac{b-\mu^*}{\sqrt{2} \sigma^*}) I_{b \in {\cal R}} \right ]
\sigma^*
\end{equation}
where $I_{a \in {\cal R}}$ ($I_{b \in {\cal R}}$) is an indicator function which equals to $1$ if $a \in {\cal R}$ ($b \in {\cal R}$) and $0$ otherwise and
\begin{equation}
\label{f}
f(x)=\frac{e^{-x^2}}{erfc(-x)}
\end{equation}

Let us now return to the multidimensional case. Since from the (truncated) conditional one-dimensional normals we no longer perform sampling but, instead, we consider their means, eq. (\ref{trunc-mean}) is altered to
\begin{equation}
\label{trunc-mean2}
\mu_i^*=\mu_i + {\mbox{\boldmath $\sigma$}}_{\neg i}^T \Sigma_{\neg i \neg i}^{-1} ({\mbox{\boldmath $w$}}_{\neg i} - {\mbox{\boldmath $\mu$}}_{\neg i}), \ \ \ i=1,\ldots,n
\end{equation}
where ${\mbox{\boldmath $w$}}_{\neg i}$ results from the current estimate of the ($n$ - dimensional) mean vector ${\mbox{\boldmath $w$}}$ of the truncated normal distribution, after removing its $i$-th coordinate.

Putting now all the previous ingredients together (that is, utilizing eqs. (\ref{cases-i-ii}), (\ref{trunc-mean2}) and (\ref{trunc-var})) we obtain the following iterative scheme

\begin{equation}
\label{new}
\begin{array}{ll}
1. & w_1^{(t)} = \mu_1^{*(t)} + \sqrt{\frac{2}{\pi}} \left [ f(\frac{\mu_1^{*(t)}-a_1}{\sqrt{2} \sigma_1^*}) I_{a_1 \in {\cal R}} - 
f(\frac{b_1-\mu_1^{*(t)}}{\sqrt{2} \sigma_1^*}) I_{b_1 \in {\cal R}} \right ] \sigma_1^* \\
2. & w_2^{(t)} = \mu_2^{*(t)} + \sqrt{\frac{2}{\pi}} \left [ f(\frac{\mu_2^{*(t)}-a_2}{\sqrt{2} \sigma_2^*}) I_{a_2 \in {\cal R}} - 
f(\frac{b_2-\mu_2^{*(t)}}{\sqrt{2} \sigma_2^*}) I_{b_2 \in {\cal R}} \right ] \sigma_2^* \\
\vdots &  \\
n. & w_n^{(t)} = \mu_n^{*(t)} + \sqrt{\frac{2}{\pi}} \left [ f(\frac{\mu_n^{*(t)}-a_n}{\sqrt{2} \sigma_n^*}) I_{a_n \in {\cal R}} - 
f(\frac{b_n-\mu_n^{*(t)}}{\sqrt{2} \sigma_n^*}) I_{b_n \in {\cal R}} \right ] \sigma_n^*
\end{array}
\end{equation}

\noindent with $\mu_i^{*(t)}$ being computed via eq. (\ref{trunc-mean2}), where ${\mbox{\boldmath $w$}}_{\neg i}^{(t)}$ (the only parameter in (\ref{trunc-mean2}) that varies through iterations) is defined as ${\mbox{\boldmath $w$}}_{\neg i}^{(t)} = [w_1^{(t)},\ldots,w_{i-1}^{(t)},w_{i+1}^{(t-1)},\ldots,w_n^{(t-1)}]^T$, that is, the most recent information about $w_i$'s is utilized. More formally, we can say that the above scheme performs sequential updating and (following the terminology used in \cite{Bert89}) it is a {\em Gauss Seidel} updating scheme.

It is reminded that, due to the type of truncation considered here (only left truncation or only right truncation per coordinate), the bracketed expression in each equation of (\ref{new}) contains {\em only one} non identically equal to zero term. In the sequel, we consider separately the cases corresponding to $a_i \in {\cal R}$ and $b_i \in {\cal R}$, i.e.,

\begin{equation}
\label{first case}
w_i^{(t)} = \mu_i^{*(t)} + \sqrt{\frac{2}{\pi}}  f(A_i^{(t)}) \sigma_i^*
\end{equation}
and 
\begin{equation}
\label{second case}
w_i^{(t)} = \mu_i^{*(t)} - \sqrt{\frac{2}{\pi}} f(B_i^{(t)})  \sigma_i^*
\end{equation}
where
\begin{equation}
\label{Ai}
A_i^{(t)}=\frac{\mu_i^{*(t)}-a_i}{\sqrt{2} \sigma_i^*}
\end{equation}
and
\begin{equation}
\label{Bi}
B_i^{(t)}=\frac{b_i-\mu_i^{*(t)}}{\sqrt{2} \sigma_i^*}
\end{equation}
$i=1,\ldots,n$, respectively,
and $f(\cdot)$ is defined as in eq. (\ref{f}).

\section{Convergence issues}

In this section we provide sufficient conditions under which the proposed scheme is proved to converge\footnote{However, it is noted that even if these conditions are slightly violated, the algorithm still works, as is verified by the experiments presented in Section 5.}.
Before we proceed, we give some propositions and remind some concepts that will be proved useful in the sequel.

\noindent {\em Proposition 1:} Assume that $\Sigma$ is a symmetric positive definite $n \times n$ matrix, ${\mbox{\boldmath $\sigma$}}_{\neg i}$ is the $i$-th column of $\Sigma$, after removing its $i$-th element and $\Sigma_{\neg i \neg i}$ results from $\Sigma$ after removing its $i$-th row and its $i$-th column. Also, let $s_{ii}$ be the $(i,i)$ element of $\Sigma^{-1}$ and ${\mbox{\boldmath $s$}}_{\neg i}^T$ be the $(n-1)$-dimensional vector resulting from the $i$-th row of $\Sigma^{-1}$ after (i) removing its $i$-th element, $s_{ii}$, and (ii) multiplying the remaining elements by $-1/s_{ii}$. Then, it holds

(i) $s_{ii}=\frac{1}{\sigma_{ii} - {\mbox{\boldmath $\sigma$}}_{\neg i}^T \Sigma_{\neg i \neg i}^{-1} {\mbox{\boldmath $\sigma$}}_{\neg i} }$ and

(ii) ${\mbox{\boldmath $s$}}_{\neg i}^T = {\mbox{\boldmath $\sigma$}}_{\neg i}^T \Sigma_{\neg i \neg i}^{-1}$.

{\vskip 10pt}
The proof of this proposition is straightforward from the inversion lemma for block partitioned matrices (\cite{Scha91}, p. 53) and the use of permutation matrices, in order to define the Schur complement $\sigma_{ii} - {\mbox{\boldmath $\sigma$}}_{\neg i}^T \Sigma_{\neg i \neg i}^{-1} {\mbox{\boldmath $\sigma$}}_{\neg i}$ for each row $i$ of $\Sigma$.
{\vskip 10pt}

\noindent {\em Proposition 2:} It is 
\begin{equation}
-\sqrt{\pi} \leq f'(x) \leq 0, \forall x \in {\cal R}
\end{equation}
where $f'(x)$ denotes the derivative of $f(x)$, which is defined in eq. (\ref{f}).

The proof of proposition 2 is given in the appendix.
{\vskip 10pt}
\noindent {\em Definition 1:} A mapping $T:X \rightarrow X$, where $X \subset {\cal R}^n$, is called {\em contraction} if for some norm $||\cdot||$ there exists some constant $\alpha \in [0,1)$ (called {\em modulus}) such that
\begin{equation}
\label{contra}
||T({\mbox{\boldmath $x$}})-T({\mbox{\boldmath $y$}})|| \leq \alpha ||{\mbox{\boldmath $x$}}-{\mbox{\boldmath $y$}}||,\ \ \forall {\mbox{\boldmath $x$}},{\mbox{\boldmath $y$}} \in X
\end{equation}

The corresponding iteration ${\mbox{\boldmath $x$}}(t+1)=T({\mbox{\boldmath $x$}}(t))$ is called {\em contracting iteration}.
{\vskip 10pt}

{\em Proposition 3 (\cite{Bert89}, pp. 182-183):} Suppose that $T:X\rightarrow X$ is a contraction with modulus $\alpha \in [0,1)$ and that $X$ is a closed subset of ${\cal R}^n$. Then

(a) The mapping $T$ has a unique fixed point ${\mbox{\boldmath $x$}}^* \in X$ \footnote{A point ${\mbox{\boldmath $x$}}^*$ is called {\em fixed point} of a mapping $T$ if it is $T({\mbox{\boldmath $x$}}^*)={\mbox{\boldmath $x$}}^*$.}.

(b) For every initial vector ${\mbox{\boldmath $x$}}(0) \in X$, the sequence $\{{\mbox{\boldmath $x$}}(t)\}$, generated by ${\mbox{\boldmath $x$}}(t+1)=T({\mbox{\boldmath $x$}}(t))$ converges to ${\mbox{\boldmath $x$}}^*$ geometrically. In particular,
$$||{\mbox{\boldmath $x$}}(t)-{\mbox{\boldmath $x$}}^*|| \leq \alpha^t ||{\mbox{\boldmath $x$}}(0)-{\mbox{\boldmath $x$}}^*||,\ \ \forall t \geq 0$$

{\vskip 10pt}
Let us define the mappings $T_i: {\cal R}^n \rightarrow {\cal R}$, $i=1,\ldots,n$ as
\begin{align}
T_i({\mbox{\boldmath $x$}}) &= \mu_i^{*}({\mbox{\boldmath $x$}}) + \nonumber \\
&+\sqrt{\frac{2}{\pi}} \left [ f(\frac{\mu_i^{*}({\mbox{\boldmath $x$}})-a_i}{\sqrt{2} \sigma_i^*}) I_{a_i \in {\cal R}} - 
f(\frac{b_i-\mu_i^{*}({\mbox{\boldmath $x$}})}{\sqrt{2} \sigma_i^*}) I_{b_i \in {\cal R}} \right ] \sigma_i^*
\label{T-i}
\end{align}
where 
\begin{equation}
\label{mu-i}
\mu_i^*({\mbox{\boldmath $x$}})=\mu_i + {\mbox{\boldmath $\sigma$}}_{\neg i}^T \Sigma_{\neg i \neg i}^{-1} ({\mbox{\boldmath $x$}}_{\neg i} - {\mbox{\boldmath $\mu$}}_{\neg i})
\end{equation}
and
\begin{equation}
\label{AB-i}
A_i({\mbox{\boldmath $x$}})=\frac{\mu_i^{*}({\mbox{\boldmath $x$}})-a_i}{\sqrt{2} \sigma_i^*},
\qquad
B_i({\mbox{\boldmath $x$}})=\frac{b_i-\mu_i^{*}({\mbox{\boldmath $x$}})}{\sqrt{2} \sigma_i^*}
\end{equation}

and the mapping $T$ as 
$$T({\mbox{\boldmath $x$}})=(T_1({\mbox{\boldmath $x$}}),\ldots,T_n({\mbox{\boldmath $x$}}) )$$

Let us define next the mapping $\hat{T}_i: {\cal R}^n \rightarrow {\cal R}^n$ as
$$\hat{T}_i({\mbox{\boldmath $x$}})=\hat{T}_i(x_1,\ldots,x_n)=(x_1,\ldots,x_{i-1},T_i({\mbox{\boldmath $x$}}),x_{i+1},\ldots,x_n)$$

Performing the sequential updating as described by eq. (\ref{new}) (one at a time and in increasing order) is equivalent to applying the mapping $S:{\cal R}^n \rightarrow {\cal R}^n$, defined as
\begin{equation}
\label{S}
S=\hat{T}_n \circ \hat{T}_{n-1} \circ \ldots \circ \hat{T}_2 \circ \hat{T}_1 
\end{equation}
where $\circ$ denotes function composition. Following the terminology given in \cite{Bert89}, $S$ is called {\em the Gauss Seidel mapping based on the mapping $T$} and the iteration ${\mbox{\boldmath $x$}}(t+1)=S({\mbox{\boldmath $x$}}(t))$ is called {\em the Gauss Seidel algorithm based on mapping $T$}.

{\vskip 10pt}
A direct consequence of \cite[Prop. 1.4, pp.186]{Bert89} is the following Proposition:
{\vskip 10pt}
{\em Proposition 4:} If $T:X\rightarrow X$ is a contraction with respect to the $l_{\infty}$ norm, then the Gauss-Seidel mapping $S$ is also a contraction (with respect to the $l_{\infty}$ norm), with the same modulus as $T$. In particular, if $X$ is closed, the sequence of the vectors generated by the Gauss-Seidel algorithm based on the mapping $T$ converges to the unique fixed point of $T$ geometrically.
{\vskip 10pt}
Having given all the necessary definitions and results, we will proceed by proving that (a) for each mapping $T_i$ it holds $|T_i({\mbox{\boldmath $x$}})-T_i({\mbox{\boldmath $y$}})| \leq ||{\mbox{\boldmath $\sigma$}}^T_{\neg i} \Sigma_{\neg i \neg i}^{-1}||_1 ||{\mbox{\boldmath $x$}} - {\mbox{\boldmath $y$}}||_{\infty}$, $\forall {\mbox{\boldmath $x$}}, {\mbox{\boldmath $y$}} \in {\cal R}^n$, where $||\cdot||_1$, $||\cdot||_{\infty}$ are the $l_1$ and $l_{\infty}$ norms, respectively, (b) if $\Sigma^{-1}$ is diagonally dominant then
$T$ is a contraction and (c) provided that $T$ is a contraction, the algorithm ${\mbox{\boldmath $x$}}(t+1)=S({\mbox{\boldmath $x$}}(t))$ converges geometrically to the unique fixed point of $T$. We remind here that the $(n-1)$-dimensional vector ${\mbox{\boldmath $\sigma$}}^T_{\neg i} \Sigma_{\neg i \neg i}^{-1}$ results from the $i$-th row of $\Sigma^{-1}$, 
exluding its $i$-th element $s_{ii}$ and dividing each element by the negative of $s_{ii}$. 

{\em Proposition 5:} For the mappings $T_i$, $i=1,\ldots,n$, it holds 
\begin{equation}
|T_i({\mbox{\boldmath $x$}})-T_i({\mbox{\boldmath $y$}})| \leq ||{\mbox{\boldmath $\sigma$}}^T_{\neg i} \Sigma_{\neg i \neg i}^{-1}||_1 ||{\mbox{\boldmath $x$}} - {\mbox{\boldmath $y$}}||_{\infty}, \ \ \forall {\mbox{\boldmath $x$}}, {\mbox{\boldmath $y$}} \in {\cal R}^n
\end{equation}

{\em Proof:} (a) We consider first the case where $a_i \in {\cal R}$. Let us consider the vectors ${\mbox{\boldmath $x$}}, {\mbox{\boldmath $y$}} \in {\cal R}^n$. Since ${\mbox{\boldmath $\mu$}}$ is constant, utilizing eq. (\ref{mu-i}) it follows that
\begin{equation}
\label{diff-mu-i}
\mu_i^*({\mbox{\boldmath $x$}})-\mu_i^*({\mbox{\boldmath $y$}}) = 
{\mbox{\boldmath $\sigma$}}^T_{\neg i} \Sigma_{\neg i \neg i}^{-1} ({\mbox{\boldmath $x$}}_{\neg i} - {\mbox{\boldmath $y$}}_{\neg i})
\end{equation}
Also, it is
\begin{equation}
\label{diff-A-i}
A_i({\mbox{\boldmath $x$}})-A_i({\mbox{\boldmath $y$}})=\frac{1}{\sqrt{2} \sigma_i^*} {\mbox{\boldmath $\sigma$}}^T_{\neg i} \Sigma_{\neg i \neg i}^{-1}({\mbox{\boldmath $x$}}_{\neg i}-{\mbox{\boldmath $y$}}_{\neg i})
\end{equation}

Taking the difference $T_i({\mbox{\boldmath $x$}})-T_i({\mbox{\boldmath $y$}})$ we have
\begin{align}
T_i({\mbox{\boldmath $x$}})-T_i({\mbox{\boldmath $y$}}) &= (\mu_i^*({\mbox{\boldmath $x$}})-\mu_i^*({\mbox{\boldmath $y$}}))+ \nonumber \\
&+ \sqrt{\frac{2}{\pi}}( f(A_i({\mbox{\boldmath $x$}}))-f(A_i({\mbox{\boldmath $y$}})) ) \sigma_i^*
\label{diff-T}
\end{align}

Since $f$ is continuous in ${\cal R}$, the mean value theorem guarantees that there exists $\xi_i \in [\min
(A_i({\mbox{\boldmath $x$}}),A_i({\mbox{\boldmath $y$}})),$ $\max(A_i({\mbox{\boldmath $x$}}), A_i({\mbox{\boldmath $y$}}))]$ such that
\begin{equation}
\label{MVT}
f(A_i({\mbox{\boldmath $x$}}))-f(A_i({\mbox{\boldmath $y$}})) = f'(\xi_i) (A_i({\mbox{\boldmath $x$}})-A_i({\mbox{\boldmath $y$}}) )
\end{equation}
Substituting eq. (\ref{MVT}) to (\ref{diff-T}) we get
\begin{align}
T_i({\mbox{\boldmath $x$}})-T_i({\mbox{\boldmath $y$}}) &= (\mu_i^*({\mbox{\boldmath $x$}})-\mu_i^*({\mbox{\boldmath $y$}}))+ \nonumber \\
&+ \sqrt{\frac{2}{\pi}}f'(\xi_i) (A_i({\mbox{\boldmath $x$}})-A_i({\mbox{\boldmath $y$}}) ) \sigma_i^*
\label{diff-T1}
\end{align}
Substituting (\ref{diff-mu-i}) and (\ref{diff-A-i}) into (\ref{diff-T1}) and after some manipulations it follows that
\begin{equation}
\label{diff-T2}
T_i({\mbox{\boldmath $x$}})-T_i({\mbox{\boldmath $y$}}) = (1+\frac{f'(\xi_i)}{\sqrt{\pi}}) 
{\mbox{\boldmath $\sigma$}}^T_{\neg i} \Sigma_{\neg i \neg i}^{-1} ({\mbox{\boldmath $x$}}_{\neg i} - {\mbox{\boldmath $y$}}_{\neg i})
\end{equation}

Taking absolute values in eq. (\ref{diff-T2}) and applying H\"{o}lder's inequality $|{\mbox{\boldmath $a$}}^T{\mbox{\boldmath $b$}}| \leq ||{\mbox{\boldmath $a$}}||_p ||{\mbox{\boldmath $b$}}||_q$, for $p=1$ and $q=\infty$, it follows that
\begin{equation}
\label{norm-T1}
|T_i({\mbox{\boldmath $x$}})-T_i({\mbox{\boldmath $y$}})| \leq ||(1+\frac{f'(\xi_i)}{\sqrt{\pi}}){\mbox{\boldmath $\sigma$}}^T_{\neg i} \Sigma_{\neg i \neg i}^{-1}||_1 ||{\mbox{\boldmath $x$}}_{\neg i} - {\mbox{\boldmath $y$}}_{\neg i}||_{\infty}
\end{equation}
Taking into account that $-\sqrt{\pi} \leq f'(\xi_i) \leq 0$ (from Proposition 2), and the (trivial) fact that
$||{\mbox{\boldmath $x$}}_{\neg i} - {\mbox{\boldmath $y$}}_{\neg i}||_{\infty} \leq ||{\mbox{\boldmath $x$}} - {\mbox{\boldmath $y$}}||_{\infty}$, it follows that
\begin{equation}
\label{norm-T}
|T_i({\mbox{\boldmath $x$}})-T_i({\mbox{\boldmath $y$}})| \leq ||{\mbox{\boldmath $\sigma$}}^T_{\neg i} \Sigma_{\neg i \neg i}^{-1}||_1 ||{\mbox{\boldmath $x$}} - {\mbox{\boldmath $y$}}||_{\infty}
\end{equation}
Thus, the claim has been proved.

(b) We consider now the case where $b_i \in {\cal R}$. Working similarly to the previous case, the difference 
$B_i({\mbox{\boldmath $x$}})-B_i({\mbox{\boldmath $y$}})$ is
\begin{equation}
\label{diff-B-i}
B_i({\mbox{\boldmath $x$}})-B_i({\mbox{\boldmath $y$}})=-\frac{1}{\sqrt{2} \sigma_i^*} {\mbox{\boldmath $\sigma$}}^T_{\neg i} \Sigma_{\neg i \neg i}^{-1}({\mbox{\boldmath $x$}}_{\neg i}-{\mbox{\boldmath $y$}}_{\neg i})
\end{equation}
while the difference $T_i({\mbox{\boldmath $x$}})-T_i({\mbox{\boldmath $y$}})$ is expressed as

\begin{align}
T_i({\mbox{\boldmath $x$}})-T_i({\mbox{\boldmath $y$}}) &= (\mu_i^*({\mbox{\boldmath $x$}})-\mu_i^*({\mbox{\boldmath $y$}})) +   \nonumber  \\
  &+ \sqrt{\frac{2}{\pi}}( f(B_i({\mbox{\boldmath $x$}}))-f(B_i({\mbox{\boldmath $y$}})) ) \sigma_i^*
\label{diff-T-B}
\end{align}
Utilizing the mean value theorem we have that there exists $\xi_i \in [\min
(B_i({\mbox{\boldmath $x$}}),B_i({\mbox{\boldmath $y$}})),$ $\max(B_i({\mbox{\boldmath $x$}}),B_i({\mbox{\boldmath $y$}}))]$ such that
\begin{equation}
\label{MVT1}
f(B_i({\mbox{\boldmath $x$}}))-f(B_i({\mbox{\boldmath $y$}})) = f'(\xi_i) (B_i({\mbox{\boldmath $x$}})-B_i({\mbox{\boldmath $y$}}) )
\end{equation}
Substituting eq. (\ref{MVT1}) to (\ref{diff-T-B}) we get

\begin{align}
T_i({\mbox{\boldmath $x$}})-T_i({\mbox{\boldmath $y$}}) &= (\mu_i^*({\mbox{\boldmath $x$}})-\mu_i^*({\mbox{\boldmath $y$}})) + \nonumber \\
 &+ \sqrt{\frac{2}{\pi}}f'(\xi_i) (B_i({\mbox{\boldmath $x$}})-B_i({\mbox{\boldmath $y$}}) ) \sigma_i^* 
\label{diff-T1-B}
\end{align}
Substituting eqs. (\ref{diff-mu-i}) and (\ref{diff-B-i}) into (\ref{diff-T1-B}), we obtain
\begin{equation}
\label{diff-T2-B}
T_i({\mbox{\boldmath $x$}})-T_i({\mbox{\boldmath $y$}}) = (1+\frac{f'(\xi_i)}{\sqrt{\pi}}) 
{\mbox{\boldmath $\sigma$}}^T_{\neg i} \Sigma_{\neg i \neg i}^{-1} ({\mbox{\boldmath $x$}}_{\neg i} - {\mbox{\boldmath $y$}}_{\neg i})
\end{equation}
From this point on, the proof is exactly the same with that of (a).    Q.E.D.
{\vskip 10pt}
{\em Proposition 6:} The mapping $T$ is a contraction in ${\cal R}^n$, with respect to the $l_{\infty}$ norm, provided that $\Sigma^{-1}$ is diagonally dominant.
{\vskip 10pt}
{\em Proof:} Let ${\mbox{\boldmath $x$}},{\mbox{\boldmath $y$}} \in {\cal R}^n$. Taking into account proposition 5, it easily follows that
$$||T({\mbox{\boldmath $x$}})-T({\mbox{\boldmath $y$}})||_{\infty} \equiv \max_{i=1,\ldots,n} \{|T_i({\mbox{\boldmath $x$}})-T_i({\mbox{\boldmath $y$}})|\} \leq$$
$$\leq \max_{i=1,\ldots,n} \{ ||{\mbox{\boldmath $\sigma$}}^T_{\neg i} \Sigma_{\neg i \neg i}^{-1}||_1 \} ||{\mbox{\boldmath $x$}} - {\mbox{\boldmath $y$}}||_{\infty} $$

Now, (a) taking into account that the $(n-1)$-dimensional vector ${\mbox{\boldmath $\sigma$}}^T_{\neg i} \Sigma_{\neg i \neg i}^{-1}$ results from the $i$-th row of $\Sigma^{-1}$, 
exluding its $i$-th element $s_{ii}$ and dividing each element by the negative of $s_{ii}$ and (b) recalling that
${\mbox{\boldmath $s$}}_{\neg i}$ is the $i$-th row of $\Sigma^{-1}$ excluding its $i$-th element $s_{ii}$, it is

$$||{\mbox{\boldmath $\sigma$}}^T_{\neg i} \Sigma_{\neg i \neg i}^{-1}||_1 =||\frac{ {\mbox{\boldmath $s$}}_{\neg i} }{s_{ii}}||_1 $$
Provided that $\Sigma^{-1}$ is diagonally dominant, it is
$$||{\mbox{\boldmath $s$}}_{\neg i}||_1 \leq |s_{ii}|,\ \ \forall i$$
or
$$||\frac{ {\mbox{\boldmath $s$}}_{\neg i} }{s_{ii}}||_1 < 1,\ \ \forall i$$
which proves the claim.   Q.E.D.

{\vskip 10pt}
{\em Theorem 1:} The algorithm ${\mbox{\boldmath $x$}}(t+1)=S({\mbox{\boldmath $x$}}(t))$ converges geometrically to the unique fixed point of $T$, provided that $\Sigma^{-1}$ is diagonally dominant.
{\vskip 5pt}
{\em Proof:} The proof is a direct consequence of the propositions 3, 4 and 6 exposed before, applied for $X={\cal R}^n$.   Q.E.D.

\section{Assessment of the accuracy of the proposed method}

An issue that naturally arises with the proposed method is how accurate the estimate of the mean is. Since it is very difficult to give a theoretical analysis of this issue, mainly due to the highly complex nature of the propsoed iterative scheme (see eq. (\ref{new})), we will try to gain some insight for this subject via experimentation. To this end, we set $\Sigma$ equal to the {\em exponential correlation} matrix, which is frequently met in various fields of applications, e.g., in signal processing applications. Its general form is
\begin{equation}
\label{exp-corr}
\Sigma_n=\left[
\begin{array}{ccccc}
1 & \rho & \rho^2 & \ldots & \rho^{n-1} \\
\rho & 1 & \rho & \ldots & \rho^{n-2} \\
\rho^2 & \rho & 1 & \ldots & \rho^{n-3} \\
\vdots & \vdots & \vdots & \ddots & \vdots \\
\rho^{n-1} & \rho^{n-2} & \rho^{n-3} & \ldots & 1
\end{array} \right], \ \ (0 \leq \rho <1)
\end{equation}
It is easy to verify that the inverse of $\Sigma_n$ is expressed as 
\begin{equation}
\label{exp-corr-1}
\Sigma_n^{-1}=\frac{1}{1-\rho^2} \left[
\begin{array}{cccccc}
1 & -\rho & 0 & \ldots & 0 & 0 \\
-\rho & 1+\rho^2 & -\rho & \ldots & 0 & 0 \\
0 & -\rho & 1+\rho^2 & \ldots & 0 & 0 \\
\vdots & \vdots & \vdots & \ddots & \vdots & \vdots \\
0 & 0 & 0 & \ldots & 1+\rho^2 & -\rho \\
0 & 0 & 0 & \ldots & -\rho & 1
\end{array} \right]
\end{equation}
Also, it is straightforward to see that $\Sigma_n^{-1}$ is diagonally dominant for all values of $\rho \in [0,\ 1)$. Thus, it is a suitable candidate for our case. In addition, it is ``controlled'' by just a single parameter ($\rho$), which facilitates the extraction of conclusions.
Note that for $\rho=0$, $\Sigma_n^{-1}$ becomes the identity matrix, while as $\rho$ increases towards $1$ the diagonal dominancy of $\Sigma_n^{-1}$ decreases (while its condition number increases). For $\rho$ close to $1$, $\Sigma_n^{-1}$ is alomost singular.

In the sequel, we consider the case of a zero mean normal distribution with covariance matrix as in (\ref{exp-corr-1}), which is truncated in the region $[a,+\infty]^n$, that is the truncation point is the same along all dimensions. Without loss of generality, this choice has been deliberately selected, in order to keep our experimental framework controlled by just two parameters, $\rho$ and $a$. Performing simulations for various combinations of the values of $\rho$ and $a$ (and for various dimensions $n$) we can gain some insight on the accuracy of the approximation of the mean provided by the proposed method. In the sequel, we use as benchmark the estimate of the mean provided by the (widely accepted as reliable) MCMC method (\cite{Robe95}).

\begin{figure*}
\center{\includegraphics[width=0.75\textwidth]{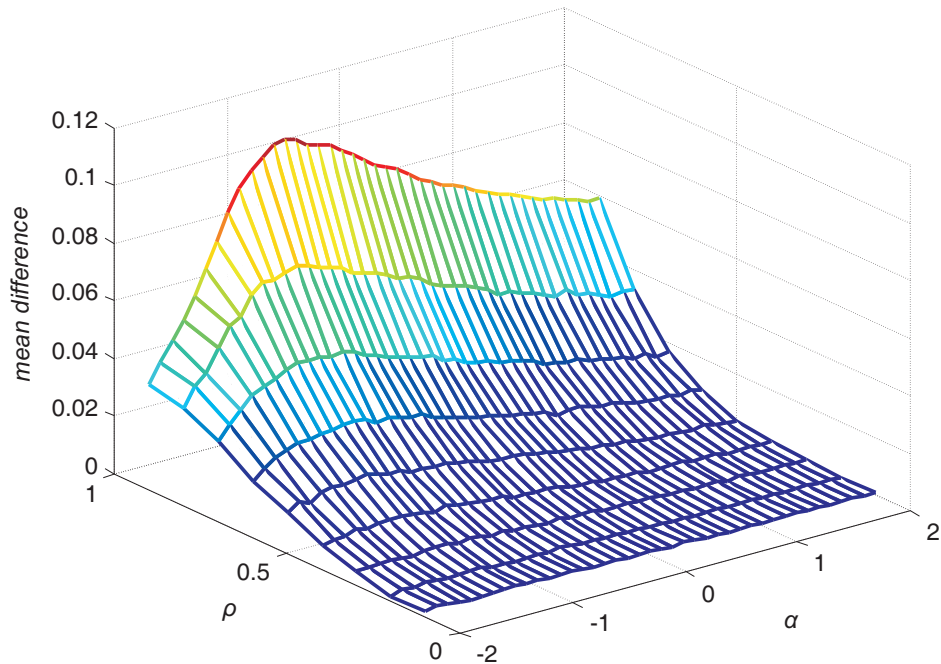}}
\caption{\small{ The mean absolute difference per coordinate between the estimates obtained by the proposed and the MCMC methods, for $n=25$ and $\rho \in (0.1,\ 0.9)$, $a \in [-2,\ 2]$.}}
\label{fig1}
\end{figure*}

Figure 1, shows a three-dimensional graph of the difference (assessed by its Euclidean norm divided by $n$) between the estimates of the truncated mean obtained by the MCMC and the proposed methods, against $\rho$ and $a$. It is worth noting that for smaller values of $\rho$ (less than $0.45$), the difference remains low (less than $0.03$), independently of the value of the cutting point $a$. On the other hand, for larger values of $\rho$ (above $0.45$), the difference increases. More specifically, it increases more for values of $a$ between (approximately) $-1$ and $1.5$. 

\begin{figure*}
\center{\includegraphics[width=0.75\textwidth]{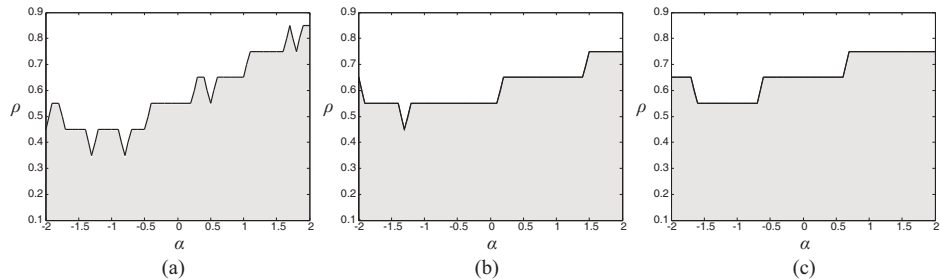}}
\caption{\small{ Contour plot for the cases where (a) $n=2$, (b) $n=25$ and (c) $n=50$. The shaded region below each contour in the figure corresponds to the case where the mean difference per coordinate between the proposed and the MCMC methods is less than $0.003$.}}
\label{fig2}
\end{figure*}

In figure 2, the shaded regions in the $(a,\rho)$ space correspond to low difference (less than $0.03$ per dimension), for $n=2,25,50$. It can be deduced that the behaviour of the proposed method is only slightly affected influenced by the dimensionality of the feature space.

As a general conclusion, one can argue that the ``more diagonally dominant'' the $\Sigma_n^{-1}$ is (i.e., the smaller the $\rho$ is), the more accurate the estimate of the mean provided by the proposed method is. From another perspective, the more $\Sigma_n$ ``approaches diagonality'' (again, as $\rho$ becomes smaller), the more accurate the obtained estimates are. The latter is also supported by the fact that in the extreme case of a diagonal covariance matrix, one has to solve $n$ independent one-dimensional problems, for which an analytic formula exists. In this case, it is easy to verify that the proposed method terminates after a single iteration (see also comments in the next section).

\section{Simulation results}

After having gained some insight on the capability of the proposed method to approximate the mean of a multivariate truncated normal distribution in the previous section, we proceed in this section with experiments where now the involved covariance matrices have no specific structure. As in the previous section, the MCMC method is used as benchmark \footnote{In the sequel, all results are rounded to the third decimal.}. 

{\em 1st experiment:} The purpose of this experiment is to compare the estimates of the mean of a truncated normal distribution obtained by the proposed and the MCMC methods, for dimensions $n$ varying from $2$ to $15$. To this end, for each dimension $n$, $k=30$ different truncated normal distributions (defined by the means and the covariance matrices of the corresponding untruncated normals, as well as their truncation points) have randomly been generated, such that, the corresponding inverse covariance matrix is diagonally dominant. For the $i$-th such distribution, $i=1,\ldots,k$, both the proposed and the MCMC methods have been applied. Letting $\mbox{\boldmath $w$}_{ni}^D$ and $\mbox{\boldmath $w$}_{ni}^{MCMC}$ denote the respective resulting estimates, the mean difference per coordinate between the two estimates is computed, i.e., $\delta_{ni}=\frac{||\mbox{\boldmath $w$}_{ni}^D-\mbox{\boldmath $w$}_{ni}^{MCMC}||_2}{n}$ and, averaging over $i$, the quantity $\Delta_n=\frac{1}{k} \sum_{i=1}^k \delta_{ni}$ is obtained. In figure 3, $\Delta_n$ is plotted versus $n$. From this figure, it can be concluded that the proposed scheme gives estimates that are very close to those given by the MCMC. Thus, the fixed point of the proposed scheme (when the diagonal dominance condition is fulfilled) can be considered as a reliable estimate of the mean of the truncated normal distribution.

\begin{figure*}
\center{\includegraphics[width=0.75\textwidth]{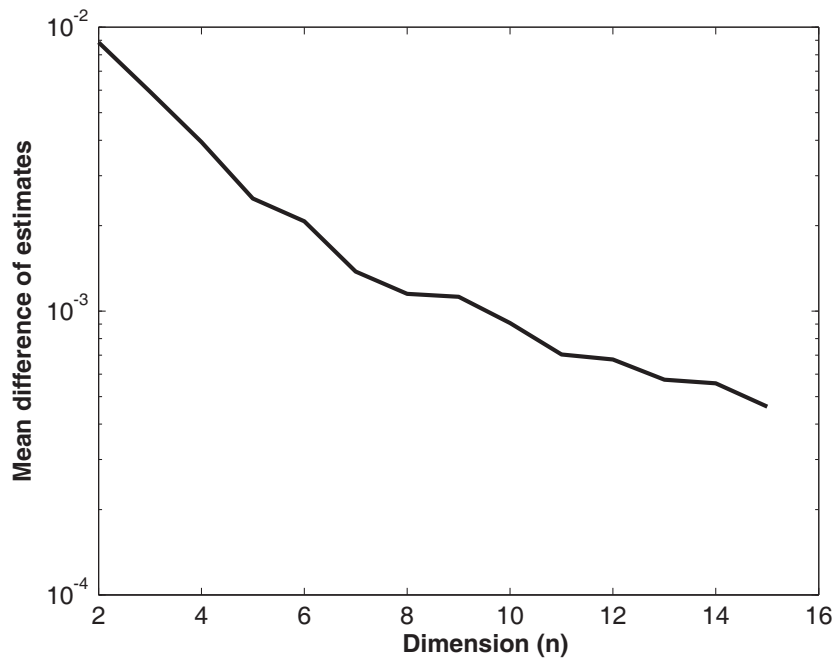}}
\caption{\small{ The mean absolute difference per coordinate between the estimates obtained by the proposed and the MCMC methods, as the dimension $n$ varies from $2$ to $15$. $30$ experiments have been conducted for each value of $n$.}}
\label{fig3}
\end{figure*}

Next, in order to show the rapid convergence of the proposed scheme against the MCMC method, we focus on a single example (however, the resulting conclusions are generally applicable). More specifically, Table 1 shows the values of the $l_1$ norm of the difference $\mbox{\boldmath $w$}^{(t)}-\mbox{\boldmath $w$}^{(t-1)}$ divided by $n$, i.e. the quantity $||\mbox{\boldmath $w$}^{(t)}-\mbox{\boldmath $w$}^{(t-1)}||_1/n$, as the number of iterations evolves, for the $5$-dimensional left truncated normal distribution defined by
$\mbox{\boldmath $\mu$}=[ 2.688,\ 9.169,\ -11.294,\ 4.311,$ $ 1.594]^T$, 
$$\Sigma=\left[ \begin{array}{ccccc}  
    $0.045$  & $-0.003$  & $0.013$ & $-0.004$ & $0.011$ \\
   $-0.003$  &  $0.056$  &  $-0.015$ & $0.008$ & $0.010$ \\
   $0.013$  &  $-0.015$  &  $0.074$ & $-0.001$ & $0.004$ \\
	 $-0.004$ &  $0.008$   &  $-0.001$ & $0.156$ & $-0.012$  \\
	 $0.011$  &  $0.010$   &  $0.004$  & $-0.012$ & $0.038$
\end{array} \right],$$
while its truncation point is 
$\mbox{\boldmath $a$}=[ 2.591,\ 8.891,\ -11.841,\ 3.353,$ $\ 0.629]^T$. It is clear that the proposed method converges rapidly to its estimate, while MCMC exhibits rather slow converge. This is the strong point of the proposed method compared to the MCMC (keeping, however, in mind the requirement for diagonal dominance of the inverse covariance matrix, for the proposed method).

\begin{table}
\centering
\begin{tabular}{rrr}
iterations & MCMC method & Proposed method \\ \hline
    $1$ & $95.130e-003$  &  $6.278e+000$ \\ 
    $2$ & $72.952e-003$  &  $571.150e-003$ \\
    $3$ & $47.163e-003$  &  $34.602e-003$ \\
    $4$ & $43.741e-003$  &  $3.897e-003$ \\
    $5$ & $35.021e-003$  &  $324.702e-006$ \\
    $6$ & $25.441e-003$  &  $27.751e-006$ \\
    $7$ & $34.047e-003$  &  $1.920e-006$ \\
    $8$ & $14.612e-003$  &  $133.025e-009$ \\
		$9$ & $17.539e-003$  &  $0$ \\
		$10$ & $21.859e-003$ &  $0$
\end{tabular}
\caption{The evolution of $||\mbox{\boldmath $w$}^{(t)}-\mbox{\boldmath $w$}^{(t-1)}||/n$ for the MCMC and the proposed methods, for the first few iterations.}
\end{table}

In the sequel we try to get an indication about the performance of the proposed method when the diagonal dominance requirement does not hold\footnote{It is reminded that no formal proof of the convergence of the proposed scheme has been given for this case.}. The following two experiments are in this direction.

{\em 2nd experiment:} We consider the three-dimensional case where 
$\mbox{\boldmath{$\mu$}}=[ 2.660,\ 9.307,\ -3.321]^T$ and 
$$\Sigma=\left[ \begin{array}{ccc}  
    $1.493$  & $-0.973$  & $-1.225$ \\
   $-0.973$  &  $4.463$  &  $3.429$ \\
   $-1.225$  &  $3.429$  &  $8.014$
\end{array} \right],$$
while the truncation point is $\mbox{\boldmath{$a$}}=[2.176,\ 8.657,\ -3.990]^T$
Note that, although in this case the inverse of the covariance matrix, which is 
$$\Sigma^{-1}=\left[ \begin{array}{ccc}  
    $0.807$  & $0.121$  & $0.072$ \\
    $0,121$  & $0.352$  & $-0.132$ \\
    $0.072$  & $-0.132$ & $0.192$
\end{array} \right],$$    
 is not diagonally dominant, the diagonal dominance condition is slightly violated (only in the third row). The estimate of the mean of the truncated normal distribution obtained by the proposed method is $\hat{\mbox{\boldmath{$\mu$}}}=[3.122,\ 10.509,\ -1.598]^T$,
which is quite close to the estimate obtained by the MCMC (the mean absolute difference per coordinate is $0.048$). It is worth noting that in this case, although eq. (\ref{norm-T}) is not satisfied ($||\max_i \{{\mbox{\boldmath $\sigma$}}^T_{\neg i} \Sigma_{\neg i \neg i}^{-1}||_1 \}>1$, since $\Sigma^{-1}$ is not diagonally dominant), the quantities $||(1+\frac{f'(\xi_i)}{\sqrt{\pi}}){\mbox{\boldmath $\sigma$}}^T_{\neg i} \Sigma_{\neg i \neg i}^{-1}||_1$, $i=1,\ldots,n$, are less than $1$ for all pairs of points $(\mbox{\boldmath $w$}^{(t)}, \mbox{\boldmath $w$}^{(t-1)})$\footnote{The dependence of this quantity from $(\mbox{\boldmath $w$}^{(t)}, \mbox{\boldmath $w$}^{(t-1)})$ is through $\xi$.}. Thus, the (tighter) bound of eq. (\ref{norm-T1}) is satisfied for all the pairs $(\mbox{\boldmath $w$}^{(t)}, \mbox{\boldmath $w$}^{(t-1)})$, which guarantees the rapid convergence of the method. However, note that, since $||(1+\frac{f'(\xi_i)}{\sqrt{\pi}}){\mbox{\boldmath $\sigma$}}^T_{\neg i} \Sigma_{\neg i \neg i}^{-1}||_1$ depends on the data (through $\xi$), it cannot be used as an upper bound in the proof of the contraction. 

{\em 3rd experiment:} We consider now the case where $\mbox{\boldmath{$\mu$}}=[ -3.968,\ -3.141,\ 8.134]^T$ and 

$$\Sigma=\left[ \begin{array}{ccc}  
    $1.082$  &  $-0.490$  &  $1.434$ \\
    $-0.490$  &  $1.088$  &  $-0.052$ \\
    $1.434$  &  $-0.052$  &  $2.711$
\end{array} \right],$$ 
while the truncation point is $\mbox{\boldmath{$a$}}=[-4.541,\ -3.358,\ 7.512]^T$.
The matrix
$$\Sigma^{-1}=\left[ \begin{array}{ccc}  
    $7.904$  &  $3.365$  &  $-4.116$ \\
    $3.365$  &  $2.352$  &  $-1.735$ \\
    $-4.116$  &  $-1.735$  &  $2.513$
\end{array} \right],$$
 is non-diagonally dominant and, in addition, the diagonal dominance condition is strongly violated (in the last two rows). We run the proposed method for several different initial conditions. It is noted that, in contrast to the 2nd experiment, the quantity $||(1+\frac{f'(\xi_i)}{\sqrt{\pi}}){\mbox{\boldmath $\sigma$}}^T_{\neg i} \Sigma_{\neg i \neg i}^{-1}||_1 ||$ is now greater than $1$ for all pairs of $(\mbox{\boldmath $w$}^{(t)}, \mbox{\boldmath $w$}^{(t-1)})$. The algorithm in all cases converges to the 
vector $\hat{\mbox{\boldmath{$\mu$}}}=[-3.859,\ -2.610,\ 8.727]^T$. 
However, in this case, the mean absolute difference per coordinate between this estimate and the one resulting from MCMC is $0.274$, which is significantly greater than that in the previous experiment.

{\em 4th experiment:} In order to exhibit the scalability properties of the proposed method with respect to the dimensionality, an additional experiment has been conducted for $n=10000$. The mean, and the covariance matrix of the corresponding normal distribution, as well as the truncation points have been selected as in the 1st experiment (note that the inverse of the covariance matrix is diagonally dominant). The algorithm gives its response in less than one minute\footnote{All experiments have been conducted on a laptop with CPU i7, 2.2GHz, 64 bit, 8GB RAM while the implementation of the method is in MATLAB.}, a time that is substantially smaller than that required from the MCMC method. 

Analysing the previous results we may draw the following conclusions:
\begin{itemize}
   \item Provided that the inverse of the covariance matrix of the untrucated normal distribution is diagonally dominant, the proposed method gives very accurate estimates of the mean of the truncated normal distribution, using as benchmark the estimates of the MCMC method.
	\item The proposed method converges much faster (in very few iterations) compared to the MCMC method.
	\item Even if the diagonal dominance constraint is slightly violated, the algorithm seems to converge to an accurate estimate of the mean of the truncated distribution, as the 2nd experiment indicates.
	\item When the diagonal dominance condition is strongly violated, the algorithm still (seems to) converge to a vector. However, this vector is a (much) less accurate estimation of the mean of the truncated normal (see the third experiment).
	\item In the special case, where $\Sigma$ (and, as a consequence $\Sigma^{-1}$) is diagonal, the matrix $\Sigma_{\neg i \neg i}^{-1}$ is (obviously) equal to zero. Thus, both eqs. (\ref{trunc-mean}) and (\ref{trunc-mean2}) implies that $\mu_i^*=\mu_i$ and eq. (\ref{trunc-var}) imply that $\sigma^*_i=\sigma_{ii}$. In other words, in this case, both the proposed and the MCMC methods solve $n$ independent one-dimensional problems, with the deterministic method giving its estimate in a single iteration, since in this case, it reduces to (\ref{case-i}) or (\ref{case-ii}). An additional observation is that the estimates provided by the two methods in this case are almost identical. Letting intuition enter into the scene and generalizing a bit, one could claim that as $\Sigma$ approaches ``diagonality'', the estimates of the mean of both methods are expected to be even closer to each other. This observation, may be an explanation of the decreasing trend observed in figure 1, since, as the dimension increases, $\Sigma$ moves closer to ``diagonality'', due to the diagonal dominance condition.
	\item Finally, the proposed method scales well with the dimensionality of the problem.
	
\end{itemize}

\section{Conclusions}
In this paper, a new iterative algorithmic scheme is proposed for the approximation of the mean value of a one-sided truncated multivariate normal distribution. The algorithm converges in very few iterations and, as a consequence, it is much faster than the MCMC based algorithm proposed in \cite{Robe95}. In addition, the algorithm is an extension of the one used in \cite{Them12}. The quality of the approximation of the mean is assessed through the case where the exponential correlation matrix is used as covariance matrix. The proof of convergence of the proposed scheme is provided for the case where $\Sigma^{-1}$ is diagonally dominant. However, experimental results indicate that, even if this condition is softly violated, the method still provides estimates of the mean of the truncated normal, however, less accurate. Finally, the method exhibits good scalablity properties with respect to the dimensinality.

\section*{Appendix}

\noindent {\em Proof of Proposition 2:}

Utilizing the fact that $\frac{d(erfc(x))}{dx}=-\frac{2e^{-x^2}}{\sqrt{\pi}}$, simple algebraic manipulations lead to
\begin{equation}
\label{der-f}
f'(x)=-2xf(x)-\frac{2}{\sqrt{\pi}} f^2(x)
\end{equation}

Before we proceed, we write down the following inequalities, which hold for $y>0$ and result from \cite[7.1.13]{Abra68}
\begin{equation}
\label{ineq}
\frac{\sqrt{\pi}}{2}(y+\sqrt{y^2+4/\pi}) \leq \frac{e^{-y^2}}{erfc(y)} < \frac{\sqrt{\pi}}{2}(y+\sqrt{y^2+2})
\end{equation}

(A) We will show first that $f'(x) \leq 0$. We separate two cases

(i) $x \geq 0$. Taking into account eq. (\ref{der-f}) and the fact that $f(x) \geq 0$, $\forall x \in {\cal R}$, 
the claim follows trivially.

(ii) $x <0$. Writing $f'(x)=-2f(x)(x+\frac{1}{\sqrt{\pi}} f(x))$, it suffices to show that 
\begin{equation}
\label{q1}
f(x) \geq -\sqrt{\pi} x
\end{equation}
Let us set $y=-x>0$. Then, $f(x)=f(-y)=\frac{e^{-y^2}}{erfc(y)}$

Taking into account the left inequality in eq. (\ref{ineq}), eq. (\ref{q1}) holds if
\begin{equation}
\label{q2}
\frac{\sqrt{\pi}}{2} (y+\sqrt{y^2+4/\pi}) \geq \sqrt{\pi} y \Leftrightarrow \sqrt{y^2+4/\pi} \geq y \Leftrightarrow 4/\pi >0
\end{equation}
Thus, the claim holds also for $x<0$.

(B) Let us consider now the following, more involved, case $f'(x) \geq -\sqrt{\pi}$. Taking into account eq. (\ref{der-f}), it suffices to prove that 
\begin{equation}
\label{q3}
\frac{2}{\sqrt{\pi}}f^2(x)+2xf(x)-\sqrt{\pi} \leq 0
\end{equation}

Let us consider the second degree polynomial of $z$
\begin{equation}
\label{poly}
\tau(z) = \frac{2}{\sqrt{\pi}}z^2+2xz-\sqrt{\pi}
\end{equation} 
Its discriminant is $D=4(x^2+2)>0$ and its two (real) roots are $z_1=\frac{\sqrt{\pi}}{2}(-x-\sqrt{x^2+2})$ and $z_2=\frac{\sqrt{\pi}}{2}(-x+\sqrt{x^2+2})$, with $z_1<z_2$. 

In order to prove (\ref{q3}) it suffices to show that $z_1 \leq f(x) \leq z_2$, or,
\begin{equation}
\label{prove}
\frac{\sqrt{\pi}}{2}(-x-\sqrt{x^2+2}) \leq f(x) \leq \frac{\sqrt{\pi}}{2}(-x+\sqrt{x^2+2})
\end{equation}
since in this case the sign of the second degree polynomial will be the opposite of that of the coefficient of $z^2$ (i.e., $\frac{2}{\sqrt{\pi}}$).
We proceed again be considering separately the cases $x \geq 0$ and $x<0$.

(i) Let $x < 0$. We set $y=-x>0$. In this case (\ref{prove}) becomes
\begin{equation}
\label{prove1}
\frac{\sqrt{\pi}}{2}(y-\sqrt{y^2+2}) \leq f(-y) \leq \frac{\sqrt{\pi}}{2}(y+\sqrt{y^2+2})
\end{equation}
where $f(-y)=\frac{e^{-y^2}}{erfc(y)}$. Taking into account (\ref{ineq}), the right hand side inequality of (\ref{prove1}) holds. In order to prove the left hand inequlity of (\ref{prove1}) (again taking into account (\ref{ineq})), it suffices to prove that

$$\frac{\sqrt{\pi}}{2}(y-\sqrt{y^2+2}) < \frac{\sqrt{\pi}}{2}(y+\sqrt{y^2+4/\pi})$$
or
\begin{equation}
-\sqrt{y^2+2} \leq \sqrt{y^2+4/\pi}
\end{equation}
which trivially holds. Thus, for $x<0$, it is $z_1 \leq f(x) \leq z_2$ and therefore $\tau(f(x)) \geq 0$. As a consequence (\ref{q3}) also holds.

(ii) Let $x \geq 0$. The left hand side inequality of (\ref{prove}) holds trivially, since in this case $z_1<0 \leq f(x)$.
We focus now on the right hand side inequality of (\ref{prove}). Taking into account that (a) $erfc(x) \leq e^{-x^2}$, for $x>0$  (see e.g. \cite{Chia03}) and (b) $erfc(x)+erfc(-x)=2$, it is
\begin{equation}
\label{q4}
\frac{e^{-x^2}}{erfc(-x)}=\frac{e^{-x^2}}{2-erfc(x)} \leq \frac{1}{2e^{x^2}-1}
\end{equation}
Combining the right hand side inequality of (\ref{ineq}) with (\ref{q4}), the right hand side inequality of (\ref{prove}) holds if 
\begin{equation}
\frac{1}{2e^{x^2}-1} \leq \frac{\sqrt{\pi}}{2}(-x+\sqrt{x^2+2})
\end{equation}
or 
\begin{equation}
\frac{1}{2e^{x^2}-1} \leq \frac{\sqrt{\pi}}{2} \frac{2}{ x+\sqrt{x^2+2} }
\end{equation}
or 
\begin{equation}
2\sqrt{\pi} e^{x^2} -\sqrt{\pi} \geq x+\sqrt{x^2+2}
\end{equation}
Since $e^z>1+z$, for $z>0$ (from the Taylor series expansion), the previous inequality holds if

$$2\sqrt{\pi} (x^2+1) -\sqrt{\pi} \geq x+\sqrt{x^2+2}$$
or
\begin{equation}
2\sqrt{\pi} x^2+\sqrt{\pi} \geq x+\sqrt{x^2+2}
\end{equation}
Since $x+\sqrt{2} \geq \sqrt{x^2+2}$, the previous inequality holds if
\begin{equation}
\label{q5}
2\sqrt{\pi} x^2-2x+\sqrt{\pi}-\sqrt{2} > 0
\end{equation}
The discriminant of the above second degree polynomial of $x$ is $D_1=4-8\sqrt{\pi}(\sqrt{\pi}-\sqrt{2})<0$. Thus, the above second degree polynomial is always positive since the coefficient of the term $x^2$ is positive. 
In other words, (\ref{q5}) holds. Therefore, the right hand side inequality of (\ref{prove}) also holds.
Thus, for $x \geq 0$, it is also $z_1 \leq f(x) \leq z_2$ and therefore $\tau(f(x)) \geq 0$. As a consequence (\ref{q3}) also holds.   Q.E.D.

{\vskip 15pt}
{\bf Acknowledgements:}
The authors would like to thank Mr. N. Malamos for his helpful suggestions in the proof of proposition 5.

\end{document}